\documentclass[a4paper, 10.5pt]{article} % Font size (can be 10pt, 11pt or 12pt) and paper size 
\usepackage{color}
\usepackage[protrusion=true,expansion=true]{microtype} % Better typography
\usepackage{graphicx} % Required for including pictures
\usepackage{wrapfig} % Allows in-line images
\usepackage{endnotes}
\usepackage[OT1]{fontenc} % Required for accented characters
\usepackage{helvet} % Specify the sans-serif font
\usepackage[margin=1.2in]{geometry}
\usepackage{placeins}
\usepackage{fancyhdr} 
\usepackage[hyperfootnotes=false]{hyperref} % Set up hyperlinks
\hypersetup{
	colorlinks   = true, % Colour links instead of ugly boxes
	urlcolor     = blue, % Colour for external hyperlinks
	linkcolor    = blue, % Colour of internal links
	citecolor   = red % Colour of citations
}
\newlength\tindent
\usepackage[normalem]{ulem}

\usepackage[symbol,hang,flushmargin]{footmisc}

\usepackage[labelfont=bf]{caption}
\usepackage{subfig}
\usepackage{longtable}

\usepackage{natbib}
\setcitestyle{authoryear,open={(},close={)}} %Citation-related commands
\newcommand{\parencite}[1]{\citep{#1}}
\newcommand{\textcite}[1]{\citet{#1}}

\makeatletter
\renewcommand\@biblabel[1]{\textbf{#1.}} % Change the square brackets for each bibliography item from '[1]' to '1.'
\renewcommand{\@listI}{\itemsep=0pt} % Reduce the space between items in the itemize and enumerate environments and the bibliography

\begin{document}
	\begin{titlepage}
		\thispagestyle{empty}
		\Large
		\noindent\rule{\textwidth}{0.4pt}
		\noindent Reserve Bank of Australia\\
		\noindent Working Paper
		\\[0.5cm]
		\textbf{Ornithologist: Towards Trustworthy ``Reasoning'' about Central Bank Communications}

		\noindent\rule{\textwidth}{0.4pt}
		\normalsize
		\noindent Disclaimer: This is a working paper, and hence it represents research in progress. Views expressed in this working paper are those of the authors and not necessarily those of the RBA.
		
		\noindent\rule{\textwidth}{0.4pt}
		\normalsize
		\noindent Dominic Zaun Eu Jones\\
		\noindent Reserve Bank of Australia\\
		\noindent 14 May 2025\\
		\noindent JonesD@rba.gov.au

	\end{titlepage}
	%\maketitle % Print the title section
	
	%----------------------------------------------------------------------------------------
	%	ABSTRACT AND KEYWORDS
	%----------------------------------------------------------------------------------------
	
	\renewcommand*{\thefootnote}{\fnsymbol{footnote}}
	%----------------------------------------------------------------------------------------
	%	TITLE & ABSTRACT
	%----------------------------------------------------------------------------------------
	
	\section*{Ornithologist: Towards Trustworthy ``Reasoning'' about Central Bank Communications\footnote{I thank the participants at the IMF, Federal Reserve Board of Governors, and the 9th Monash-Paris-Warwick-Zurich-CEPR Text-As-Data Workshop for their valuable comments.}}
	\renewcommand*{\thefootnote}{\arabic{footnote}}
	\setcounter{footnote}{0}
	
	\noindent \textit{I develop Ornithologist, a weakly-supervised textual classification system and measure the hawkishness and dovishness of central bank text. Ornithologist uses ``taxonomy-guided reasoning'', guiding a large language model with human-authored decision trees. This increases the transparency and explainability of the system and makes it accessible to non-experts. It also reduces hallucination risk. Since it requires less supervision than traditional classification systems, it can more easily be applied to other problems or sources of text (e.g. news) without much modification. Ornithologist measurements of hawkishness and dovishness of RBA communication carry information about the future of the cash rate path and of market expectations.}
	\\\\
	\noindent \textbf{Keywords}: Central Bank Communication, Large Language Models, Natural Language Processing, Weakly Supervised Text Classification, Taxonomy Construction
	
	\noindent \textbf{JEL Codes}: E52, E58, C38, C88
	
	%----------------------------------------------------------------------------------------
	%	ANALYTICAL NOTE BODY
	%----------------------------------------------------------------------------------------
	
	\subsection*{Motivation}
	
	%------------------------------------------------
	
There are few sources of text as scrutinised as central bank communication – at least in financial markets. Central bankers themselves are probably the chief scrutineers. Extracting signal from any kind of text, in many cases, requires a textual classifier. In policy contexts, these classifiers should ideally be performant and explainable. In this paper, I develop \textit{Ornithologist}, a weakly-supervised textual classification scheme built around explainability. To validate the approach, I measure the hawkish and dovish tilt of our communications.

All market participants are interested in the hawkish or dovish tilt of central banks. Central banks care because it is how they communicate\footnote{It might be more correct to say it is a \textit{framing} of how we communicate. That said, central banks do have to make decisions, and mapping the sentiment (whatever that is) of our communication to these decisions is, essentially, the hawk-dove frame.} their view of the future state of the economy, or their reaction function, or however you'd like to phrase it. Members of the public care because macroeconomic policy affects them. Market participants care because a lot of money is at stake. So much so that they develop automated systems to measure hawkishness and dovishness \parencite{lupton_welcome_2023}.

Traditional fully supervised methods are expensive, often requiring significant domain-expert effort to label a large enough set of examples. By combining large language models (LLMs) and a human-constructed taxonomy of economic concepts and reasoning guides I significantly reduce the supervision signal required. Due to this the method is more easily applied to other measurement problems, like \textcite{romer_does_1989} narrative (textual) monetary policy shocks – though I do not attempt these in this paper.

In fact, the narrative approach that \textcite{romer_presidential_2023} summarise\footnote{In Table 1 of their paper.} guides this work. It is worthwhile repeating here. A rigorous narrative (textual) analysis requires:
\begin{itemize}
	\item ``A reliable narrative source''. I focus on RBA statements and minutes, which are consistent, detailed, and accurate, albeit not real-time.
	\item ``A clear idea of what one is looking for''. More traditional supervised classification techniques can miss context, and simply prompting a LLM is, frankly, too magical to be trustworthy. I use a technique called \textit{taxonomy-guided reasoning} to use LLMs in a more epistemically acceptable way.
	\item ``[Approaching] the source dispassionately and consistently''. If nothing else, LLMs certainly have this to recommend them.
	\item ``[Documenting] the narrative evidence carefully''. Unlike traditional approaches, the output of this system can explain its reasoning and be checked.
\end{itemize}

Unlike \textcite{romer_does_1989}, we do not need to manually read all the input text.

\subsection*{Summary of contribution}
\textit{Ornithologist} uses a hand constructed economic taxonomy paired with a set of human-authored decision trees to guide a LLM in making classifications. To reduce hallucination risk, it uses constrained generation techniques to limit the LLM to follow a decision tree exactly. The main contribution it makes is to show how one can construct a classification system based on explanation while making precisely how those explanations are formed transparent, verifiable, and modifiable by non-experts. That is: it is accessible so to foster understanding of and trust in the system. To the best of my knowledge, a system with these properties is novel in central banking contexts. I also provide a taxonomy useful for the Australian economic context.

While the measurement of hawkish and dovish sentiment is not novel, I find that Ornithologist measurements carry information about the future of the cash rate path and are an effective summary of RBA forecasts in an estimate of our policy reaction function.
	
\subsection*{Measuring things with text}
For a review of textual approaches in economics, see \textcite{gentzkow_text_2019}. \textcite{lucca_measuring_2009} use word-level sentiment\footnote{Technically, ``semantic orientation'' scores, which is essentially a trained dictionary sentiment model.} to measure hawkishness and dovishness. \textcite{tobback_between_2017} extend the dictionary approach by training a classifier that considers words together, rather than adding up scores. Dictionary-based approaches, while simple, can be biased because the mere presence or absence of a term misses the complexity of language. Negations or other modifiers, for example, might not be fully controlled for.\footnote{That said, dictionary methods are often quite useful.}

More modern large language models address the shortcomings of dictionary approaches. \textcite{hansen_can_2024} and \textcite{peskoff_gpt_2023} demonstrate that ChatGPT can measure hawkishness and dovishness. JP Morgan publishes a Hawk-Dove Score \parencite{lupton_welcome_2023} uses BERT-based models – a different (and smaller) family of LLMs to autoregressive (GPT-style) ones.
 
\textcite{Gambacorta2024} use BERT-based models fine-tuned to ``understand'' central bank text. They then use these models to classify hawkishness or dovishness using data annotated by \textcite{gorodnichenko_voice_2023}. They report accuracies of around 84 per cent at the sentence level, outperforming autoregressive LLMs. The result agrees with \textcite{zhong_can_2023}: ``there is still a performance gap between ChatGPT and [BERT-based models] in terms of average performance.'' That said, fine-tuning strategies applied to autoregressive LLMs (or using more sophisticated ones) does boost performance to parity. BERT-based models are also more resource-efficient than autoregressive models.

The amount of context models have access to matters. Annotating hawkishness and dovishness at the sentence level can be difficult, due to information boundaries. Pronouns, for example, may be missing their antecedents, though semantic chunking\footnote{Grouping sentences together based on how similar they are (under some definition: usually cosine similarity between embeddings).} may serve to alleviate this. The fundamental difficulty of the task (e.g. should we make a subtextual logical leap?) can also be the source of disagreement between expert annotators. This fundamental and reasonable disagreement can cause classifier accuracies to be biased upwards \parencite{gordon_disagreement_2021}.

\subsection*{If BERT is so good, why are you using autoregressive LLMs?}
For Ornithologist, I use off-the-shelf autoregressive models. If, per \textcite{Gambacorta2024}, BERT-based models are just as performant as autoregressive models on this task, and they're more efficient, why use autoregressive models?

\paragraph{Off-the-shelf flexibility} Open-weight autoregressive models are more flexible across multiple tasks and types of data without requiring fine-tuning.

\paragraph{Explainability (of reasoning)} While BERT-based models can achieve state-of-the-art performance, they do not provide reasoning. Autoregressive LLMs can provide something like reasoning – there are now entire categories of ``reasoning" models. This sort of reasoning, however, can be epistemically problematic in the policy domain. By using human-authored decision trees, we convert free-form chain-of-thought reasoning into expert-verified reasoning pathways.

\paragraph{Ease of modification} BERT-based classifiers need to be retrained on new data if concepts or domains change. Ornithologist needs less retraining. Little modification would be required to measure on economic news rather than central bank communication.

\vspace{10pt}

While the first reason is primarily about effort, all reasons work together to increase trust in the system. Classifier output alone – using any architecture – does not provide interpretability, maintainability, and understandability. That said, it is worth investigating whether reformatting Ornithologist to use BERT-based models can improve efficiency.

\subsection*{LLM reasoning}
The ability of a large language model to solve problems seems correlated with the amount of output space it has. The more space, the more it can generate explanations, justifications, options, and so on, which amounts to reasoning (or something like it). \textcite{wei_chain--thought_2024} introduce chain of thought, a technique that engineers prompts for LLMs to produce exactly this kind of sequential reasoning. \textcite{kojima_large_2022} show that LLMs are so-called ``zero-shot" reasoners: by prompting it to ``think step by step'', the LLM can generate its own chains of thought without prompt engineering. \textcite{yao_tree_2024} introduce tree of thoughts, where an LLM generates an expanding tree of intermediate thoughts and evaluates which are the most promising paths to follow. There are now many LLMs using these approaches – so-called ``reasoning models''. If we wanted a model to be able to explain itself, why not use one of these?

For a general reasoning system – we probably should use them. However, we know the domain ahead of time: hawkish-dovish sentiment. We can ``bake in'' reasoning pathways: human experts can author decision trees for the LLM to follow. Human authoring has several advantages. Most importantly, the reasoning steps (i.e. questions) are more likely to be trusted and be of high quality, having come from an expert human source. Second, novel situations can be quickly adapted to by aligning with expert scenario analysis. Third, it forces us to be clear about our assessments. Finally, trees can be elicited from many experts and ensembled together, perhaps approximating the wisdom of crowds. Ultimately, the trade-off we are making by baking in decision trees is verifiability and trust instead of general-purpose prowess.

\subsection*{Approach}
\begin{figure}[h!]
\begin{center}
	\caption{\textbf{Ornithologist system}}
	\label{fig:ornithologist-system}
	\includegraphics[width=0.8\textwidth]{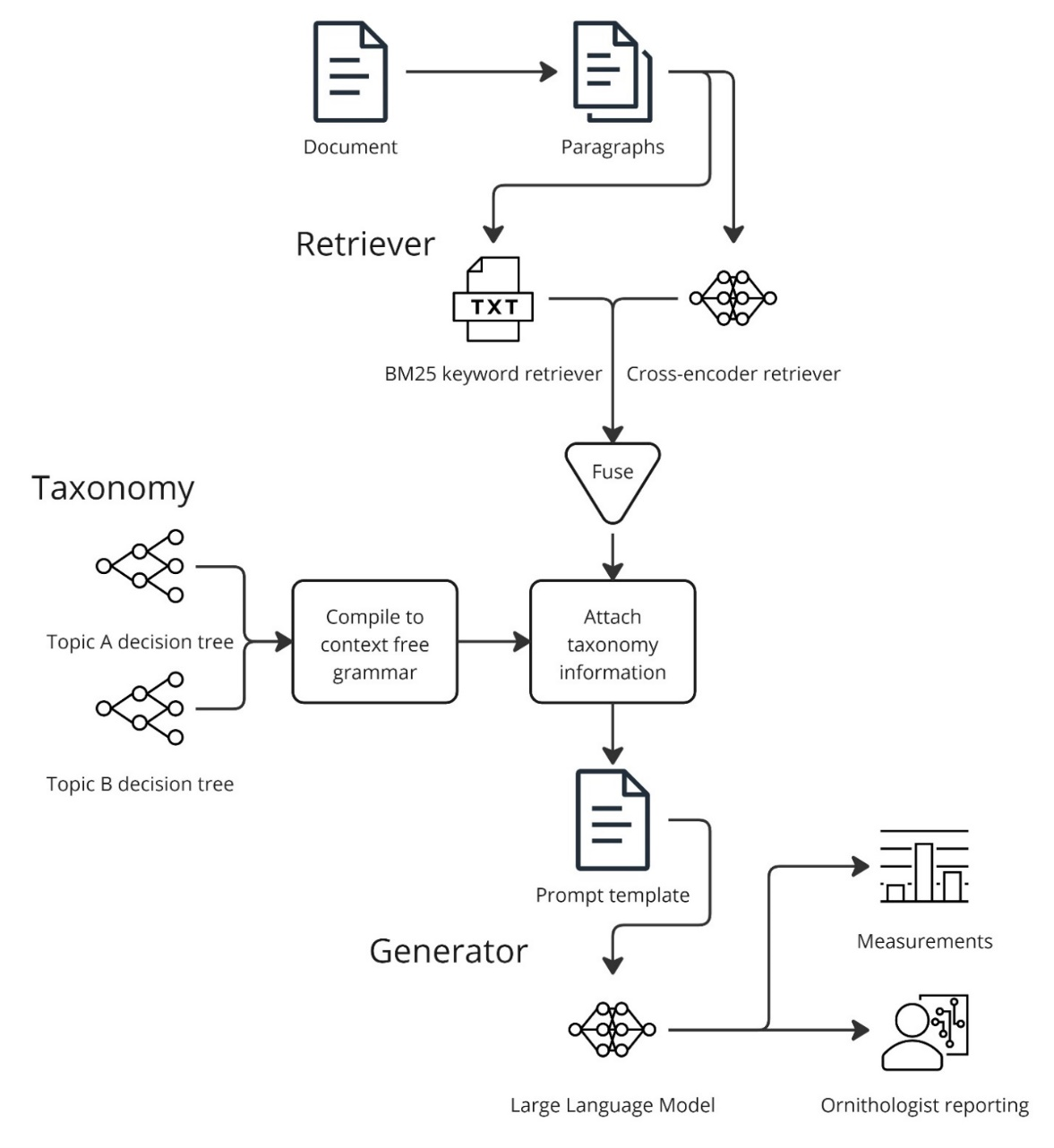}
\end{center}
\end{figure}

Ornithologist has three components: a human-authored taxonomy, a retriever, and a generator\footnote{Similar to a Retrieval-Augmented Generator (RAG) system, where an LLM's knowledge gaps are augmented by retrieving relevant content from some knowledge base.}. Figure \ref{fig:ornithologist-system} summarises the system. The retriever is a text matching model that annotates passages of text with relevant topics. These topics align with decision trees and other information in the taxonomy. The generator takes these topics, a prompt, converts the decision trees into a context-free grammar which constrains generation, and generates output. In this case, it delivers paragraph- and sentence-level classifications for the hawkishness or dovishness of central bank communication. By representing decision trees as a grammar and forcing the generator to sample from it, I can guarantee that the system will follow human direction.

\subsubsection*{Human-authored taxonomy}
I define 66 topics that are relevant for central bank communication. These topics are further related to higher-level themes (e.g. ``core mandate'', ``external sector") and lower-level phrases (e.g. ``prices and wages") in a taxonomy. Each topic is associated with a hand-authored decision tree. For example, the ``labour market, extensive margin" topic tree asks about whether the labour market indicators point towards slack or tightness and whether inflation is described as a risk. The terminals of these decision trees contain human-authored assessments.

For a full list of topics and an example decision tree, see Appendix \ref{appendix-taxonomy}. A sample: Forecasting, Monetary Policy, Fiscal Policy, Inflation Target, Labour Market (extensive and intensive margins), Financial Crises, Residential Real Estate, Equities Markets, Consumer and Business Confidence, Financial Risks, Currencies, Mining, International Monetary Policy, Demographics, and Saving.

I defined the taxonomy and topics inductively. First, I ran AutoPhrase, a phrasal segmentation algorithm, on all available RBA minutes \parencite{liu_mining_2015,shang_automated_2018}. Phrasal segmentation algorithms break text into high-quality ``units of meaning": ``full employment", for example, should be considered one unified concept. This resulted in roughly 2500 phrases. I manually read these and came up with an initial taxonomy.  For each phrase longer than one word I ask OpenAI's \texttt{gpt-4o-mini} to annotate which of the topics each phrase belongs to. I then manually verified each annotation.

\subsubsection*{Retriever}
Given a passage of text (e.g. a sentence or paragraph), the retriever matches relevant topics. I adopt a hybrid approach, using both traditional keyword relevance algorithms and more modern transformer-based approaches. The rankings returned by these approaches are then merged via reciprocal rank fusion.

The keyword-based retriever uses BM25 \parencite{robertson_probabilistic_2009}. I match the input document with each (manually verified) phrase in the taxonomy. The parent topics of the top-k phrases (in this case, 10), are then weighted by how many times they appear, resulting in a ranking.

The transformer-based retriever uses a ``cross-encoder" model based on RoBERTa \parencite{reimers_sentence-bert:_2019,liu_roberta:_2019,sanh_distilbert_2019}. The model matches a passage of text with however many topics are relevant, using both the passage itself and a ``surface" (i.e. human-understandable) representation of the topic. The surface representations include both the name of the topic and examples: e.g. ``Labour market, extensive (e.g. unemployment rates, participation, etc.)". To train it, I use the keyword-based retriever to retrieve the top-k most relevant sentences from the RBA minutes corpus for each phrase. These are then tagged as (confidently) belonging to however many topics the phrase is associated with. I then ask OpenAI's \texttt{gpt-4o} to produce 5 paraphrases of each sentence. For each sentence, negatives (i.e. topics that aren't relevant) are sampled uniformly at random. This results in 2,295,420 sentence-topic pairs. On a 2 per cent validation sample, the system achieves an F1\footnote{Micro. An F1 is a measure of retrieval quality, or predictive performance. It is the harmonic mean of precision (how many retrieved topics are correct?) and recall (how many correct topics were retrieved?). In particular} score of 91 per cent.

The hybrid retriever annotates on the paragraph level – even though the cross-encoder is trained on the sentence level. I found that using the model on paragraphs more reliably extracted the most relevant topics and reduced noise. For example, a sentence may mention tax policy, but it is not core to the paragraph. There are other ways to smooth this out: taking topics that appear in at least some fraction of sentences, for example. In the end, I used the model on entire paragraphs for simplicity.

\subsubsection*{Generator}
The generator takes paragraph annotations, follows the decision tree for each annotated topic, and replies with paragraph-level classification, sentence-level classifications, and ``reasoning". I adopt similar classifications to \textcite{hansen_can_2024} and \textcite{lupton_welcome_2023}: dovish, leaning dovish, neutral, leaning hawkish, and hawkish.

I use Meta's \texttt{Llama 3.2 3B} and Microsoft's \texttt{Phi 3.5 mini} LLMs quantised\footnote{With full precision, each parameter is represented by a 16-bit floating point number. By quantising this number to 8 or fewer bits, we can fit larger models into memory and speed up generation, albeit with a quality cost.} to 8 bits as the generator. It is worth noting that these are not reasoning models. These small LLMs can be run internally on Bank hardware. This means we can use Ornithologist on data we are not comfortable sending over the wire.

A question in the decision tree only has a finite number of answers. For example, the question ``Is inflation described as a risk, are policymakers willing to tolerate inflationary pressures, or is there no mention at all?"  has three possible answers: ``inflation risk discussed", ``willing to tolerate inflation pressures", and ``no mention of inflation". The generator is forced to choose from one of the possible answers by sampling first from a \textit{context-free grammar} (CFG). I compile each decision tree to a CFG which is provided to the LLM along with the prompt.\footnote{I use llama.cpp as the backend, which provides this capability.} The LLM, constrained to generate only valid text under the grammar, must answer within the possible set. The LLM may answer incorrectly – especially if the question is malformed, not appropriate, or is missing the correct answer. But the grammar guarantees that it will not go ``off script". This reduces the risk that the LLM will generate too ``magical" or nonsensical output, while giving it latitude to provide reasoning wherever appropriate.

Ornithologist considers entire paragraphs instead of single sentences. It could use the entire document: it is limited by the context window of the underlying LLM. By parametrising the prompts or the guiding taxonomy, more relevant context could be brought in. One current limitation of the system is that some decision trees require an assessment of whether inflation is above or below target. While the model can infer this in many cases, the state of inflation is observable. By changing the prompt or decision trees based on the current state of inflation, or unemployment, or whatever, we can provide any sort of context to the model we please. This is similar to ``multi-modal" models where one might provide an image. Instead, we are considering economic timeseries data. I leave this to future work.

\subsection*{Validation and results}

\subsubsection*{Classifier validation}
After I ran the system on all available RBA minutes, I sampled 100 paragraphs uniformly at random to manually validate. \ref{tab:validation-results} reports the results.

% \usepackage{graphicx}
\begin{table}[]
	\caption{\textbf{Validation results}}
	\label{tab:validation-results}
	\resizebox{\textwidth}{!}{%
		\begin{tabular}{rcccccc}
			\multicolumn{1}{l}{} & \multicolumn{3}{c}{\textbf{Three-class}}               & \multicolumn{3}{c}{\textbf{Five-class}}                \\
			\multicolumn{1}{l}{} & \% (Pts)              & \% (Pts)            & \%       & \% (Pts)              & \% (Pts)            & \%       \\
			Level                & Acc/Mean Err. (Llama) & Acc/Mean Err. (Phi) & Baseline & Acc/Mean Err. (Llama) & Acc/Mean Err. (Phi) & Baseline \\
			Paragraph            & 65 (0.06)             & 54 (0.14)           & 38       & 58 (0.19)             & 43.4 (0.18)         & 31       \\
			Sentence             & 67.4 (0.06)           & 53 (0.08)           & 37.3     & 51 (0.16)             & 39 (0.15)           & 34      
		\end{tabular}%
	}
\end{table}

I report results for both five classes (including leaning categories) and three classes (only dovish, neutral, and hawkish) at both paragraph- and sentence-level. These accuracies are not necessarily comparable with previously reported accuracies considering the context window and assessment differences: to manually validate, I read the entire paragraph and attempt to follow the decision trees to come to a decision. I also report baseline percentages: these represent the accuracy of a classifier just picking the modal human annotation.

I report accuracies and mean errors, rather than RMSEs. Mean errors are positive, indicating that the system is scoring sentences consistently more hawkish than I am: although, not by much. Note the range (in points) is -1 to 1 for the three-class case and from 1 to 5 for the five-class.

For Llama, in general, paragraph-level classifications seem more accurate (of 60-65 per cent), though the sentence-level three-class result is slightly higher. Since the model reports classifications at both levels, we could use validation results like these to decide how much weight to put on one or the other (at validation, it was 100 per cent at the sentence level, which is generally consistent with the paragraph).

Five-class results are significantly lower at the sentence level, which is the largest difference between these results and others. Partly, this reflects that the model was not specifically designed for this task: the reasoning takes place at the paragraph-level. It also reflects that this system is weakly supervised rather than fully so. A final source of differences are my own personal failings as an annotator and taxonomy builder.

There is a significant difference between Phi and Llama: Phi seems to underperform in accuracy (though performing similarly using mean errors) – particularly for the five-class problem. Qualitatively, however, Phi output is similar to Llama. This could reflect the difficulty of the underlying annotation problem – or model instability due to quantisation.

\subsubsection*{Economic validation}

\begin{figure}[h!]
	\centering
	\vspace{-5pt}
	\parbox{0.47\textwidth}{
		\caption{}
		\vspace{-10pt}
		\includegraphics[width=0.45\textwidth]{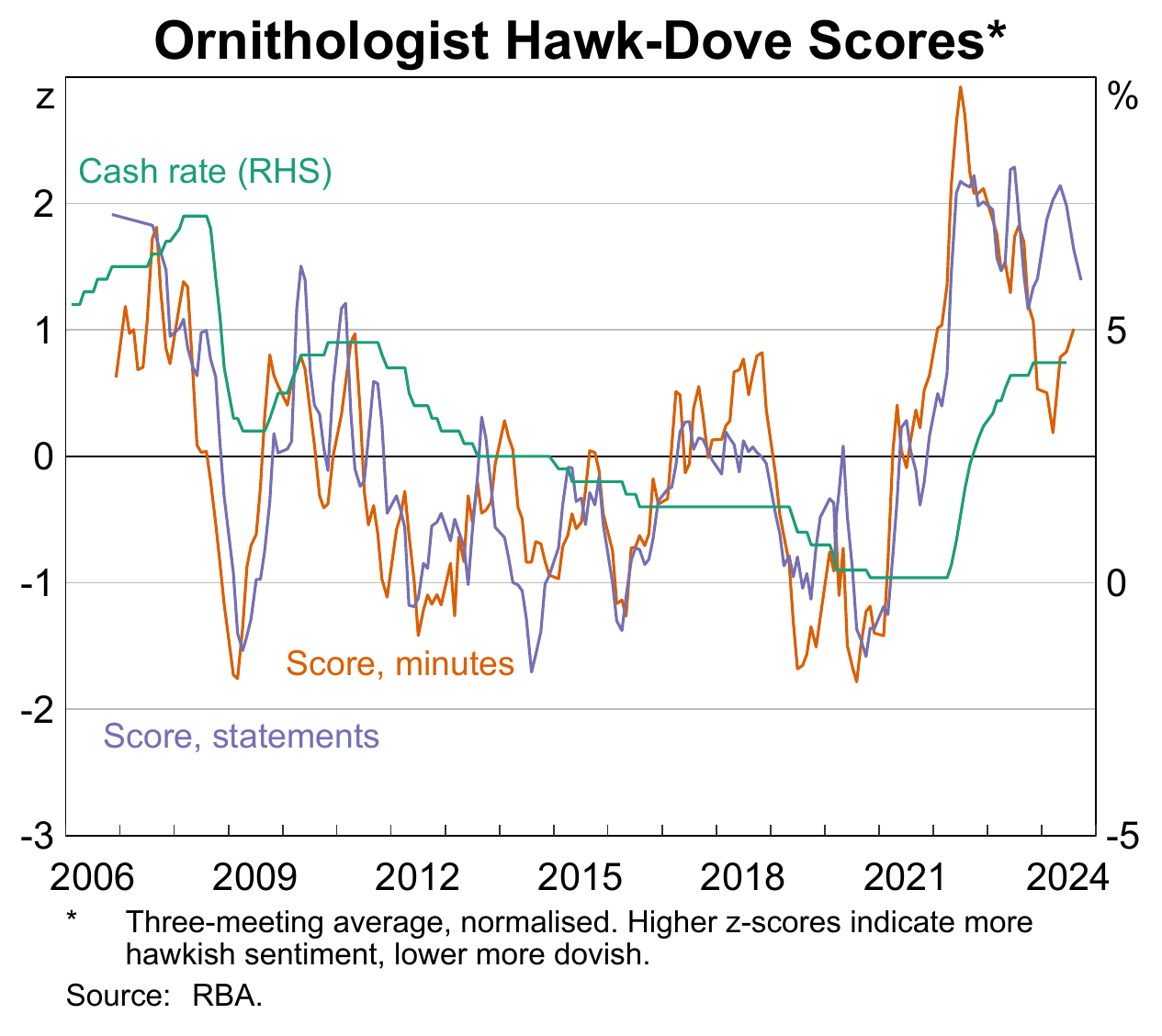}
		\label{fig:hds-timeseries}}
	\qquad
	\begin{minipage}{0.47\textwidth}
		\caption{}
		\vspace{-10pt}
		\includegraphics[width=1\textwidth]{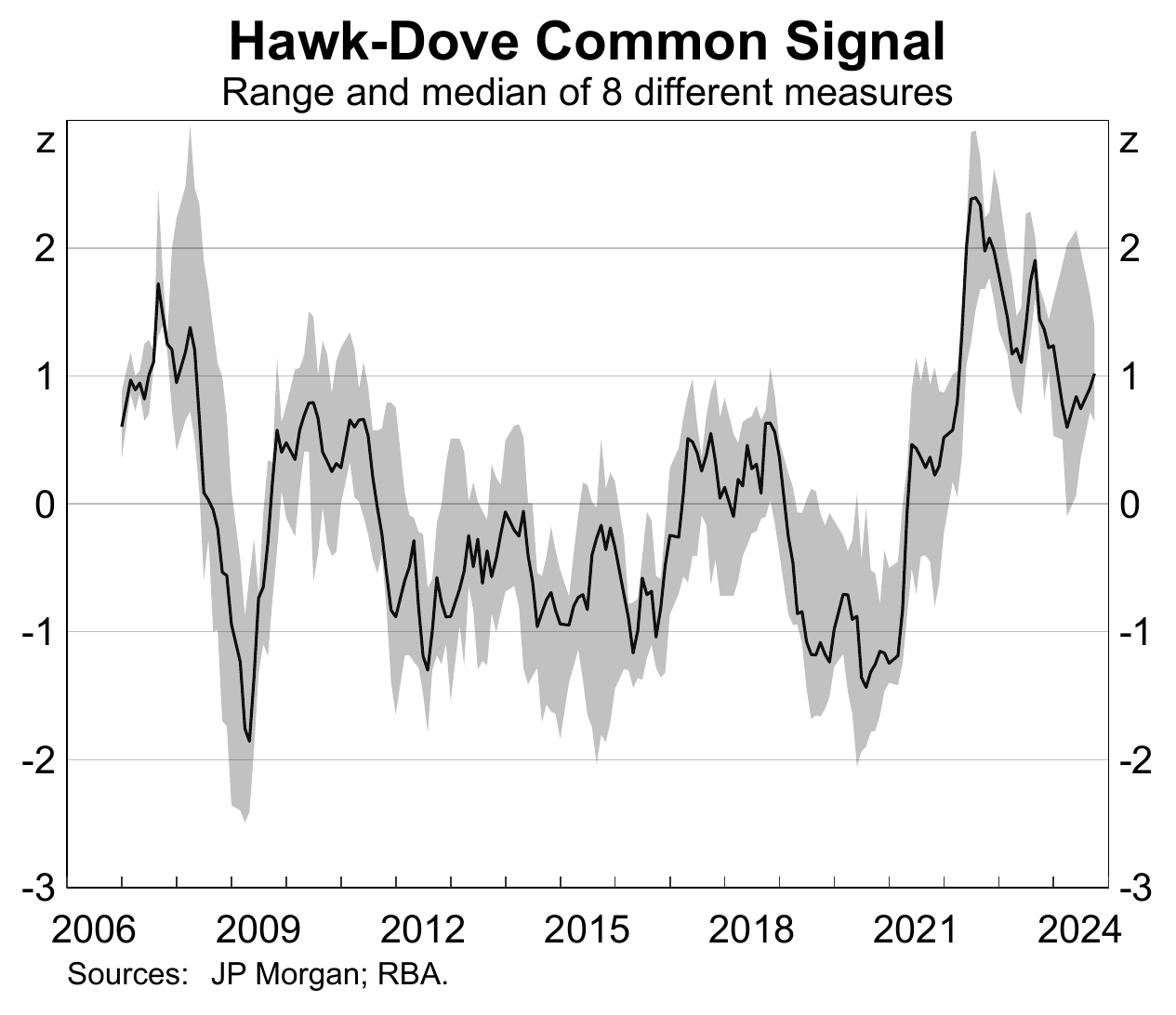}
		\label{fig:hds-common-signal}
	\end{minipage}
\end{figure}
	
For each document – from RBA statements and minutes – I compute sentence-level classifications from Ornithologist and take an average. Figure \ref{fig:hds-timeseries} shows the normalised score for each document in the corpus. There is a clear low-frequency correlation between the cash rate and Ornithologist's measure of hawkishness and dovishness. While the measurement is noisy – I plot three-meeting moving averages – they appear to lead movements in the cash rate (and indeed, the scores Granger-cause the cash rate).

\begin{wrapfigure}{r}{0.5\textwidth} % Inline image example
	\begin{center}
		\vspace{-15pt}
		\caption{}
		\vspace{-10pt}
		\label{fig:correlation-matrix}
		\includegraphics[width=0.5\textwidth]{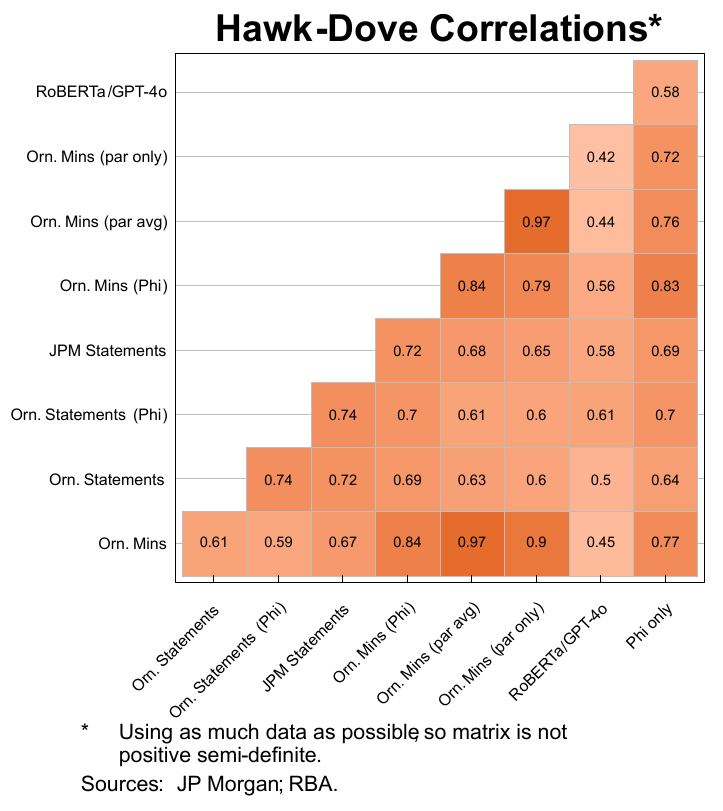}
	\end{center}
\end{wrapfigure} 

The series also appears to have high convergent validity. Convergent validity asks whether different measures of the same concept agree. If they don't, it seems likely that a mistake has been made or the concept is ill-defined. There are several other ways to measure hawkishness and dovishness – and several other ways to parametrise Ornithologist. JP Morgan's published Hawk-Dove Score is useful for comparison. I also include a method fine-tuning RoBERTa on annotations from \texttt{gpt-4o mini}, and a prompt-only approach with a local LLM. Figure \ref{fig:hds-common-signal} shows the range and median of these different measures of hawkishness and dovishness. In general, they are all highly correlated with each other and carry a common signal, even though they may be computed via totally different approaches. This increases my confidence that the measurements from Ornithologist are reasonably valid. Figure \ref{fig:correlation-matrix} shows the correlation matrix between the different measures: all show moderately high to very high correlations with each other.

Ornithologist reports classifications at the paragraph and sentence level. I use the sentence-level classifications, as they are informed by the paragraph-level classification but also admit variance within. Admittedly, it's not immediately clear what the best choice is, but using either (or an average between the two) makes little difference.

How much information about policy do the hawk-dove scores carry? Following \textcite{fadda_central_2022} I estimate a policy reaction function (except with an ordered logit instead of a probit – though results are similar). Table \ref{tab:reg-prf} shows the results. The response variable is monetary policy stance: tightening, loosening, or no change. It is categorical: increases in the cash rate map to the tightening category. I include changes to the Term Funding Facility (TFF) and yield curve control policies.\footnote{This should make the zero lower bound problem less of a problem: the Bank could have eased policy further under this definition if it wanted. For details on the TFF, please see \href{https://www.rba.gov.au/mkt-operations/term-funding-facility/}{this}, and for yield curve control, \href{https://www.rba.gov.au/monetary-policy/reviews/yield-target/index.html}{here}.} That is, even if the cash rate doesn't move, if TFF or yield curve control policies were announced, these change the policy stance. The dependent variables include Hawk-Dove scores for minutes and statements: both the previous and second-last. I also include one-year-ahead year-ended growth forecasts for CPI, GDP, and the unemployment rate. When the minutes and statements are included, the forecasts are no longer significant. This might indicate that the minutes are both an effective summary of information relevant to monetary policy, and an important tool to communicate our reaction function. The sample period covers 2006-2007 (depending on availability) through 2024.

Following this, a natural question to ask is around the connection with market expectations. While a proper answer is out-of-scope for this note, I perform a similar exercise as above using OIS-implied cash rate expectations.\footnote{Thanks to Dmitry Titkov for providing the code to compute these.} Instead of discretising these, I use the difference between the expected cash rate (as at the board date) across 1-, 3-, 6- and 12-month tenors and the current cash rate. Using the same variables as above neatly sidesteps the issue of the minutes coming out 2 weeks later. Table \ref{tab:reg-ois} shows the results.

These regressions ask if we can use forecasts and Ornithologist to predict market expectations of cash rate changes across several tenors. Note the lower AICs for all specifications with Ornithologist measurements. These regressions are admittedly unsophisticated, but highlight that Ornithologist could be useful for understanding how our communications might affect expectations. 

% \usepackage{graphicx}
\begin{table}[]
	\caption{\textbf{Hawk-Dove measurements carry information about policy}}
	\label{tab:reg-prf}
	\resizebox{\textwidth}{!}{%
		\begin{tabular}{rccccccccc}
			\multicolumn{1}{l}{} & \multicolumn{3}{r}{\textbf{Baseline}}                                                                      & \multicolumn{3}{r}{\textbf{With communications   (Phi)}}                                                   & \multicolumn{3}{r}{\textbf{With communications   (Llama)}}                                                \\ \hline
			\multicolumn{1}{l}{} & \textbf{Estimate}                                               & \textbf{OR*} & \textbf{$p$}                & \textbf{Estimate}                                               & \textbf{OR*} & \textbf{$p$}                & \textbf{Estimate}                                              & \textbf{OR*} & \textbf{$p$}                \\ \hline
			CPI F'cast           & \textbf{\begin{tabular}[c]{@{}c@{}}1.07\\ (0.22)\end{tabular}}  & 2.92         & \textbf{\textless{}0.001} & \begin{tabular}[c]{@{}c@{}}-0.26\\ (0.33)\end{tabular}          & 0.77         & 0.43                      & \begin{tabular}[c]{@{}c@{}}0.26\\ (0.29)\end{tabular}          & 1.3          & 0.37                      \\
			Activity F'cast      & \begin{tabular}[c]{@{}c@{}}-0.24\\ (0.19)\end{tabular}          & 0.78         & 0.21                      & \begin{tabular}[c]{@{}c@{}}-0.24\\ (0.23)\end{tabular}          & 0.79         & 0.29                      & \begin{tabular}[c]{@{}c@{}}-0.29\\ (0.24)\end{tabular}         & 0.75         & 0.22                      \\
			Unemp F'cast         & \textbf{\begin{tabular}[c]{@{}c@{}}-0.04\\ (0.02)\end{tabular}} & 0.96         & \textbf{0.06}             & \begin{tabular}[c]{@{}c@{}}0.00\\ (0.02)\end{tabular}           & 1.00         & 0.99                      & \begin{tabular}[c]{@{}c@{}}0.01\\ (0.02)\end{tabular}          & 1.01         & 0.78                      \\
			$HDS_{mins,t-1}$         &                                                                 &              &                           & \textbf{\begin{tabular}[c]{@{}c@{}}2.91\\ (0.81)\end{tabular}}  & 18.4         & \textbf{\textless{}0.001} & \textbf{\begin{tabular}[c]{@{}c@{}}4.23\\ (0.89)\end{tabular}} & 69.0         & \textbf{\textless{}0.001} \\
			$HDS_{mins,t-2}$         &                                                                 &              &                           & \textbf{\begin{tabular}[c]{@{}c@{}}3.16\\ (0.85)\end{tabular}}  & 23.5         & \textbf{\textless{}0.001} & \textbf{\begin{tabular}[c]{@{}c@{}}1.80\\ (0.85)\end{tabular}} & 6.03         & \textbf{0.04}             \\
			$HDS_{statement,t-1}$    &                                                                 &              &                           & \textbf{\begin{tabular}[c]{@{}c@{}}1.64\\ (0.51)\end{tabular}}  & 5.17         & \textbf{\textless{}0.01}  & \begin{tabular}[c]{@{}c@{}}-0.06\\ (0.49)\end{tabular}         & 0.94         & 0.91                      \\
			$HDS_{statement,t-2}$    &                                                                 &              &                           & \textbf{\begin{tabular}[c]{@{}c@{}}-1.88\\ (0.52)\end{tabular}} & 0.15         & \textbf{\textless{}0.001} & \begin{tabular}[c]{@{}c@{}}0.19\\ (0.49)\end{tabular}          & 1.21         & 0.70                      \\
			Accuracy (Baseline)  & \multicolumn{3}{c}{73.1\% (73.6\%)}                                                                        & \multicolumn{3}{c}{79.2\% (73.2\%)}                                                                        & \multicolumn{3}{c}{76\% (73.2\%)}                                                                         \\
			AIC                  & \multicolumn{3}{c}{278.55}                                                                                 & \multicolumn{3}{c}{205.04}                                                                                 & \multicolumn{3}{c}{209.49}                                                                                \\
			Observations         & \multicolumn{3}{c}{197}                                                                                    & \multicolumn{3}{c}{183}                                                                                    & \multicolumn{3}{c}{183}                                                                                  
		\end{tabular}%
	}
	
	\vspace{5pt}
	
	{\raggedright \small * Odds ratios.\\Ordered logistic regressions. Estimates significant at the 10 per cent level are bolded. Standard errors are reported in brackets.\\Source: RBA.}
\end{table}

% \usepackage{graphicx}
\begin{table}[]
	\caption{\textbf{Relationship with market expectations}}
	\label{tab:reg-ois}
	\resizebox{\textwidth}{!}{%
		\begin{tabular}{rcccccccc}
			\multicolumn{1}{c}{\textbf{(Tenor)}} & \multicolumn{2}{c}{\textbf{1-month}} & \multicolumn{2}{c}{\textbf{3-month}} & \multicolumn{2}{c}{\textbf{6-month}} & \multicolumn{2}{c}{\textbf{12-month}} \\ \hline
			\multicolumn{1}{l}{}                 & \textbf{Baseline}  & \textbf{Comms}  & \textbf{Baseline}  & \textbf{Comms}  & \textbf{Baseline}  & \textbf{Comms}  & \textbf{Baseline}   & \textbf{Comms}  \\ \hline
			CPI F'cast                           & \textbf{0.08}      & -0.08           & \textbf{0.14}      & \textbf{-0.15}  & \textbf{0.2}       & -0.16           & 0.13                & -0.11           \\
			Activity F'cast                      & -0.06              & -0.05           & -0.09              & -0.08           & -0.1               & -0.09           & -0.06               & -0.02           \\
			Unemp F'cast                         & \textbf{-0.01}     & -0.01           & \textbf{-0.02}     & -0.01           & \textbf{-0.02}     & -0.01           & \textbf{-0.02}      & 0.0             \\
			$HDS_{mins,t-1}$                         &                    & \textbf{0.36}   &                    & \textbf{0.67}   &                    & \textbf{1.0}    &                     & \textbf{1.27}   \\
			$HDS_{mins,t-2}$                         &                    & \textbf{0.21}   &                    & \textbf{0.39}   &                    & \textbf{0.54}   &                     & \textbf{0.41}   \\
			$HDS_{statement,t-1}$                    &                    & \textbf{0.11}   &                    & \textbf{0.17}   &                    & 0.17            &                     & 0.08            \\
			$HDS_{statement,t-2}$                    &                    & \textbf{-0.14}  &                    & \textbf{-0.21}  &                    & \textbf{-0.35}  &                     & \textbf{-0.49}  \\ \hline
			AIC                                  & 66.58              & 7.96            & 271.09             & 184.34          & 395.31             & 302.67          & 457.6               & 385.32         
		\end{tabular}%
	}
	
	\vspace{5pt}
	
	{\raggedright \small OLS with robust (HC1) standard errors. Estimates significant at the 5 per cent level are bolded. I use the Phi HDS estimates. \\ Sources: Bloomberg; RBA}
\end{table}

\subsubsection*{Reporting}
\begin{figure}[h!]
	\begin{center}
		\caption{\textbf{Ornithologist reporting}}
		\label{fig:ornithologist-reporting}
		\includegraphics[width=\textwidth]{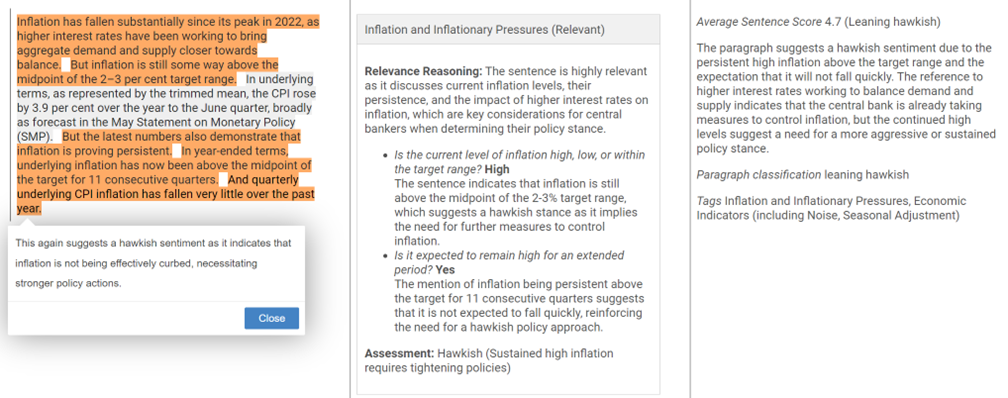}
	\end{center}
\end{figure}

Ornithologist also includes an interrogative tool for inspecting results. Figure \ref{fig:ornithologist-reporting} shows sample output for one paragraph from the September 2024 decision statement. On the left column, the paragraph is reported with each sentence highlighted according to the model. A user may click on each sentence to drill down into the reasoning behind a classification – note the tooltip pointing to the last sentence. The centre column contains the taxonomy guided reasoning report. It reports the decision tree results, including intermediate reasoning, for each relevant taxonomy topic. The rightmost column reports the overall synthesis and paragraph classification and some auxiliary information.

The reports can be generated on-demand for any type of central bank text and are distributed as self-contained HTML files.

Finally, Ornithologist can also report document-level summaries of hawkish and dovish sentences. This assists users in understanding why, at a high level, a document is hawkish or dovish, and is especially useful to understand the sentiment's evolution. Take, for example, the August and September 2024 decision statements. I take the set of sentences classified as hawkish (or dovish) from both sentences and rearrange it into a ``similar points" set and a ``new points" set. I use a sentence transformer model \parencite{reimers_sentence-bert:_2019} trained for semantic similarity to do so: sentences that are similar enough go in one set, and the others form the new points. Table \ref{tab:narratives} presents an extract.

\begin{table}[h!]
	\centering
	\caption{\textbf{The evolution of hawkish language from August to September 2024}}
	\label{tab:narratives}
	\begin{tabular}{p{0.45\linewidth} | p{0.45\linewidth}}
		\begin{tabular}[c]{@{}l@{}}(Selected) similar sentences\\(to August   2024 statement)\end{tabular}                                                                  & \begin{tabular}[c]{@{}l@{}}New sentences\\(in September 2024 statement)\end{tabular}                                \\ \hline
		Inflation   has fallen substantially since its peak in 2022, as higher interest rates   have been working to bring aggregate demand and supply closer towards   balance. & But   inflation is still some way above the midpoint of the 2–3 per cent target   range.                                 \\
		The   economic outlook is uncertain and recent data have demonstrated that the   process of returning inflation to target has been slow and bumpy.                       & But   the latest numbers also demonstrate that inflation is proving persistent.                                          \\
		Some   central banks have eased policy, although they remain alert to the risk of   persistent inflation.                                                                & In   year-ended terms, underlying inflation has now been above the midpoint of the   target for 11 consecutive quarters. \\
		Policy   will need to be sufficiently restrictive until the Board is confident that   inflation is moving sustainably towards the target range.                          & Geopolitical   uncertainties remain elevated, which may have implications for supply chains.                            
	\end{tabular}
\end{table}

We can then use a LLM to summarise the differences (or simply read them if short enough):

Similar: ``Inflation has decreased since its peak in 2022, but remains above the target range of 2-3\%. The economic outlook is uncertain, and there are risks that inflation could rise further. Wages growth has peaked but is still above sustainable levels, contributing to inflation concerns. The central bank remains committed to returning inflation to target, despite the challenges and uncertainties in the economy."

New: ``The input sentences discuss the current state of inflation in the economy, highlighting that it remains persistent and above target levels. Despite some easing of monetary policy, inflation is expected to remain high for some time, and central banks are remaining vigilant to potential risks. The labour market remains tight, and geopolitical uncertainties may impact supply chains, further complicating the inflation outlook."

These summaries can be overlaid on timeseries charts.

\subsection*{Discussion}

\subsubsection*{Error analysis}
\textbf{Retrieval failures}. Sometimes, the retriever does not match the most appropriate topic. If this occurs, Ornithologist will miss important contextual knowledge which will likely not be corrected by the paragraph summary and synthesis step. The economic disruption topic, for example, seems to be a particular weakness for the retriever (though the overall scoring doesn't seem bothered by this in the COVID period).

\textbf{Broad taxonomy limitations}. The human-designed taxonomy can cause errors. For example, I do not have a China topic, which meant the questions the model asked and answered were not appropriate for much of the China-related text. Instead, it asked about trade and currency movements (which are not unrelated, but perhaps a logical step too far).

\textbf{Poorly worded questions}. Since the LLM is restricted to answer from a finite set, if the question doesn't match up nicely this can cause errors. This usually happens when the most appropriate answer is ``none of the above", but you haven't allowed the model to answer along those lines.

\textbf{LLM errors}. Sometimes the LLM simply gets a question wrong, even though it has all the context it needs. This may be due to my use of relatively tiny LLMs: \texttt{Llama 3.2} and \texttt{Phi 3.5 mini} have roughly 3-3.5 billion parameters. State-of-the-art LLMs have upwards of hundreds of billions of parameters or more.

\textbf{Multiple topics can help in avoiding errors}. In some cases, retrieving multiple similar topics (e.g. forecasting, which includes questions about GDP, and GDP itself) can help the model avoid or reduce the severity of errors. Diversity of information helps; this may motivate a more advanced sampling-based retrieval scheme.

\textbf{Trees could benefit from numerical parametrisation}. In some cases, a decision tree asks about where inflation is relative to target. Many times, the model is correct, but usually because it is making an inference: logical leaps are impressive, but best avoided where possible. Sometimes, the model is incorrect because it picks a different number. In general, LLMs are not good with numbers, so we should avoid burdening them by parametrising our decision trees.

\textbf{Modifiability of decision trees is useful}. During model development, errors could be fixed by adjusting the decision trees. For example, adding a new subtree or a more appropriate answer, or rewording a particularly difficult question. This allows a user without technical skills to change and improve the system.

\textbf{Hawk-Dove assessment is a fundamentally hard problem}. Perhaps unsurprising in retrospect, there were many paragraphs that did not lend themselves to an obvious classification. Many phenomena (i.e. topics) have uncertain effects on inflation. One might filter them out. However, a reasonable response is that if it wasn't somehow relevant to policy, it wouldn't have been written. Ultimately, this makes the manual validation harder to perform and interpret alone, but the downstream validation exercises provide strong arguments that Ornithologist is measuring what it is meant to be.

\subsubsection*{Some thoughts about central bank communication}
\textbf{Subtext matters}. One of the reasons that central bank communication is more comprehensible to economists is that the text often assumes an underlying mental model. The document may state something – about, say, commodity prices – and expect that the reader will connect this to inflation without the mechanisms needing to be explicitly stated. This is subtext: something unstated that requires relatively advanced inferential capacity. In Ornithologist, this is represented explicitly as the human-authored taxonomy. For communicators, making more of these implicit logical links explicit – moving subtext into text – would aid both human and machine comprehension.

\textbf{Context matters}. Often there is information at the paragraph, document, or economic context level that is required to understand a paragraph. Systems that can incorporate different modes of context: whether textual or numeric (perhaps visual?), can aid understanding.

\textbf{Communication is an important source of information about the reaction function}. The regression results above – though meant for validation purposes – demonstrate that one can learn a lot about a central bank's reaction function given its communications. This relies on (a) communication effectively summarising all relevant information for a decision (e.g. the forecasts – which our communication seems to do) and (b) being effectively measurable: this highlights the need for clear communication. If the goal was for the wider public to understand our reaction function better, this motivates work on comprehensibility: e.g. reducing subtextual cues or conceptual complexity \parencite{McMahon2023}.

\textbf{Readability}. Ornithologist, and a person, is more likely to make a mistake the more complex a sentence or paragraph becomes. Since Ornithologist (and LLMs more generally) are automated, it might be useful as an effective measure of comprehensibility. From the above: subtextual cues, missing context, conceptual complexity, and concept density (e.g. talking about many things in one sentence or paragraph) would presumably be captured by this approach.

\subsubsection*{Limitations}
\textbf{Resources}. Being based on LLMs, Ornithologist has relatively high resource requirements. While, presumably, this will improve, computing full back-histories for large amounts of text (e.g. international comparisons) or comparing different taxonomies requires some time.

\textbf{Model complexity}. While LLMs provide reasoning as an output, the models themselves are black boxes. It's also not clear how to select the best LLM, given the complexity of the validation task.

\textbf{Retriever re-training is sometimes required}. While the taxonomy trees and other auxiliary information can be changed on-the-fly, if a new category is added the retriever needs to be re-trained. This isn't fatal: it can be augmented with rules (e.g. for scenario analysis) or the cross-encoder retriever can be turned off. That said, this is the main factor slowing down taxonomy iteration.

\subsection*{Conclusion and future work}

Ornithologist is a weakly-supervised text classification system. It combines taxonomy-guided reasoning with state-of-the-art LLMs to provide trustworthy classifications with only moderate supervision requirements. Taxonomy-guided reasoning relies on expert-constructed topical decision trees to guide how a LLM ``reasons". When applied to measuring the hawkishness and dovishness of RBA communication, it passes manual validation checks, convergent validity checks, and is informative for monetary policy stance and market expectations. Information can be presented and summarised from the sentence-, paragraph-, document-, or even between-document levels. I develop reports for these purposes.

Future work includes: parametrising the taxonomy decision trees with the current state of the economy (e.g. CPI and the unemployment rate), building a framework to use Ornithologist as a scenario analysis tool, and applying the method to estimate a series of \textcite{romer_does_1989} narrative monetary policy shocks for Australia.
\\\\
	Dominic Zaun Eu Jones \\
	Data Science Hub \\
	Economic Research Department \\
	14 May 2025
\\\\
Circulated internally 6 December 2024, updated 14 May 2025.

\pagebreak
%\printbibliography
\bibliography{ornithologist}
\pagebreak

\appendix
\section{Economic taxonomy}
\label{appendix-taxonomy}
\subsection{Full taxonomy}

\small
\begin{longtable}{p{0.38\linewidth} | p{0.58\linewidth}}
		\textbf{Mnemonic}          & \textbf{Topic}                                                                                             \\ \hline
		EC-FORECAST                & Forecasting (including Consensus Economics, Scenarios, etc.)                                               \\
		EC-INDICATOR               & Economic Indicators (including Noise, Seasonal Adjustment)                                                 \\
		POL-MONETARY               & Monetary Policy                                                                                            \\
		POL-STIMULUS               & Economic Stimulus (incl. Government Grants, Incentives, Rebates,   Subsidies)                              \\
		POL-FISCAL                 & Fiscal Policy (incl. austerity)                                                                            \\
		POL-GOVBUDGET              & Government Budget (Deficits and Surpluses)                                                                 \\
		POL-TAX                    & Taxation                                                                                                   \\
		POL-LEGISLATIONREGULATION  & Legislation and Regulation                                                                                 \\
		CORE-INFLATION             & Inflation and Inflationary Pressures                                                                       \\
		CORE-INFLATIONEXPECTATIONS & Inflation Expectations                                                                                     \\
		CORE-TARGET                & Inflation Target (2-3 per cent)                                                                            \\
		CORE-PRODUCTIVITY          & Productivity                                                                                               \\
		CORE-CAPACITY              & Productive Capacity (e.g. aggregate demand/supply, capacity)                                               \\
		CORE-LABOUREXTENSIVE       & Labour Market, Extensive Margin (e.g. Employment, Unemployment,   Participation, Hires and Quits, Layoffs) \\
		CORE-LABOURINTENSIVE       & Labour Market, Intensive Margin (e.g. Hours Worked,   Underutilisation, Part vs Full Time)                 \\
		CORE-LABOURCAPACITY        & Labour Market Capacity (e.g. labour market slack or tightness,   the NAIRU)                                \\
		CORE-SKILLS                & Worker Skills and Human Capital                                                                            \\
		CORE-WAGES                 & Wages, Salaries, and Employee Compensation (including Enterprise   Bargaining)                             \\
		CORE-ACTIVITY              & GDP and Economic Activity (e.g. Domestic Demand, National Accounts)                                        \\
		CORE-SUPPLYSHOCKS          & Supply Shocks                                                                                              \\
		CORE-DEMANDSHOCKS          & Demand Shocks                                                                                              \\
		CORE-DISRUPTION            & Economic Disruption (e.g. trade tensions, COVID-19 pandemic)                                               \\
		CORE-BUSACTIVITY           & Business Activity (e.g. Profits and Solvency, Inventories)                                                 \\
		CORE-CYCLES                & Economic Cycles (Recessions and Expansions)                                                                \\
		CORE-FINDISRUPTION         & Financial Crises                                                                                           \\
		CORE-HOUSEHOLDINCOMES      & Household Incomes and Budgets                                                                              \\
		CORE-WEALTH                & Wealth                                                                                                     \\
		CORE-COMPETITION           & Competition                                                                                                \\
		CORE-INVESTMENT            & Investment and Capital Expenditure                                                                         \\
		CORE-CONSUMPTION           & Consumption                                                                                                \\
		CORE-TRADABLENONTRADEABLE  & Tradable and Non-Tradable Sectors                                                                          \\
		CORE-MANUF                 & Manufacturing Sector                                                                                       \\
		CORE-SERVICES              & Services Sector                                                                                            \\
		CORE-AFFORDABILITY         & Affordability and Cost of Living                                                                           \\
		CORE-CAPITALSTOCK          & Capital Stock (incl. infrastructure)                                                                       \\
		RE-COMMERCIAL              & Commercial Real Estate                                                                                     \\
		RE-RESIDENTIAL             & Residential Real Estate (e.g. Housing, Rents, Dwelling   Construction and Investment)                      \\
		RE-CONSTRUCTION            & Building Approvals and Construction                                                                        \\
		CREDIT-INTERESTRATES       & The official cash rate or other interest rates                                                             \\
		CREDIT-VOLATILITY          & Financial volatility                                                                                       \\
		CREDIT-BANKING             & Banking Sector (including Resolution and Macroprudential   Policies)                                       \\
		CREDIT-CREDITGROWTH        & Credit Growth and Allocation                                                                               \\
		CREDIT-PRICING             & Credit and Asset Pricing (e.g. Asset Prices, Yield Curves)                                                 \\
		CREDIT-EQUITIES            & Equities Markets                                                                                           \\
		CREDIT-BONDS               & Bond Markets and Securitisation (e.g. RMBS, ABS)                                                           \\
		CREDIT-HOUSEHOLDDEBT       & Household Debt (e.g. Mortgages, Credit Cards)                                                              \\
		CREDIT-CORPDEBT            & Corporate Debt (e.g. Corporate Bonds, Business Loans)                                                      \\
		CREDIT-GOVTDEBT            & Government Debt (e.g. Australian Government Securities, Treasury   Bills)                                  \\
		CREDIT-INFRASTRUCTURE      & Financial Market Infrastructure (e.g. central counterparties,   stock exchanges)                           \\
		RISK-CONFIDENCE            & Consumer and Business Confidence                                                                           \\
		RISK-FINRISK               & Financial risks (e.g. credit risk, duration risk, risk-on or   risk-off sentiment)                         \\
		RISK-GEOPOLITICAL          & Geopolitical risk (e.g. war, terrorist attacks)                                                            \\
		RISK-INSURANCE             & Insurance                                                                                                  \\
		EXT-CURRENCIES             & Currencies                                                                                                 \\
		EXT-INTLECON               & International Economics and Capital Flows                                                                  \\
		EXT-TRADE                  & Trade (e.g. Imports and Exports)                                                                           \\
		EXT-MINING                 & Mining and Resources Sector Activity                                                                       \\
		EXT-COMMODITIES            & Oil and Bulk Commodity Markets (e.g. Oil, Gas, Iron Ore)                                                   \\
		EXT-AGRICULTURAL           & Agricultural Commodities                                                                                   \\
		EXT-INTLMONETARYPOLICY     & International Monetary Policy Comparisons                                                                  \\
		FUN-DEMOGRAPHICS           & Demographics and Population (including International Students   and Migration)                             \\
		FUN-CLIMATE                & Weather Events and Environmental Policies                                                                  \\
		SAV-SAVING                 & Saving                                                                                                     \\
		SAV-SUPER                  & Superannuation Schemes                                                                                     \\
		OTH-CBGOVERNANCE           & Central Bank Governance                                                                                    \\
		OTH-ORGANISATIONS          & International Organisations (e.g. IMF, Chinese Communist Party,   ASEAN)                                  
		\label{tab:full-taxonomy}
\end{longtable}

\normalsize

\subsection{Example decision tree}

\begin{figure}[h!]
	\begin{center}
		\caption{}
		\label{fig:decision-tree}
		\includegraphics[width=\textwidth]{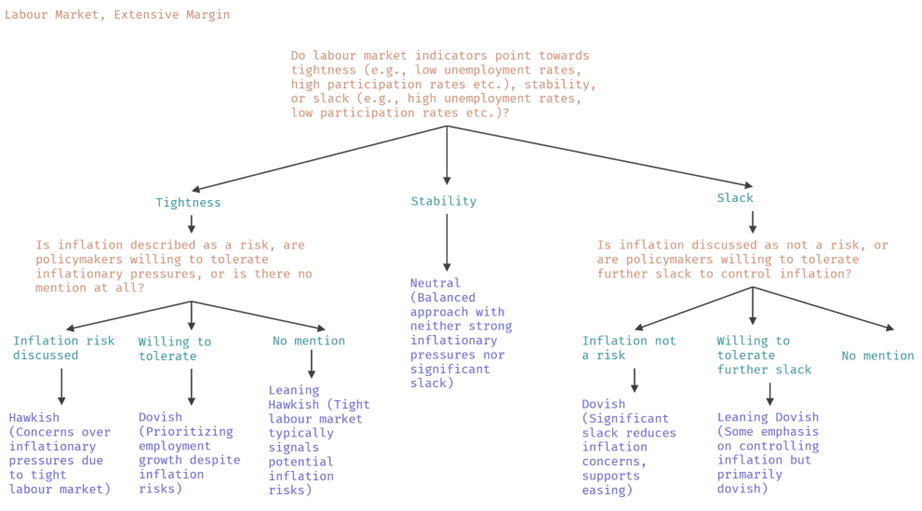}
	\end{center}
\end{figure}
	
\end{document}